\documentclass{PoS}

\usepackage{amsfonts}
\usepackage{amsmath}
\usepackage{subfigure}
\usepackage{tipa}
\usepackage{arydshln}
\graphicspath{{./figs/}}

\newcommand{\ol}{ \overline }
\newcommand{\wtd}[1] { \widetilde{#1} }

\newcommand{\GeV}{\mathop{\rm GeV}\nolimits}

\newcommand{\tr}{\textrm{tr}}

\newcommand{\bk}{$B_K\thickspace$}
\newcommand{\ek}{$\varepsilon_K\thickspace$}
\newcommand{\MSb}{\mathop{\rm \overline{MS}}\nolimits}
\newcommand{\tree}{\text{tree}}
\newcommand{\I}{\mathbf{I}}
\newcommand{\II}{\mathbf{II}}
\title{Non-perturbative Renormalization of Four-Fermion Operators
  Relevant to \bk with Staggered Quarks}

\ShortTitle{NPR of \bk}

\author{
Hwancheol Jeong, 
\speaker{Jangho Kim},
Jongjeong Kim, 
Weonjong Lee, 
Jeonghwan Pak, 
Sungwoo Park
\\
Lattice Gauge Theory Research Center, CTP, and FPRD, \\
Department of Physics and Astronomy, \\
Seoul National University, Seoul, 151-747, South Korea\\
E-mail: \email{wlee@snu.ac.kr}}

\author{SWME Collaboration}

\abstract{
We present preliminary results of matching factors of the four-fermion
operators relevant to \bk, which are obtained using the
non-perturbative renormalization (NPR) method in the RI-MOM scheme
with HYP-smeared improved staggered fermions.
We use the MILC asqtad coarse ($a \cong 0.12\,$fm) ensembles with $20^3
\times 64$ geometry and $am_{\ell}/am_s = 0.01/0.05$.
We compare NPR results with those of one-loop perturbative matching.
}

\FullConference{The 32nd International Symposium on Lattice Field Theory,\\
		23-28 June, 2014\\
		Columbia University New York, NY}

\begin{document}

\section{Introduction}
The indirect CP violation parameter, \ek in the neutral kaon system is
very well known with $\approx 0.5\,\%$ precision from experiments
\cite{Agashe:2014kda}.
Our theoretical estimate of \ek directly from the standard model (SM)
has 3.4$\sigma$ tension with the experiment in the exclusive $V_{cb}$
channel \cite{Lee:2014bjl}.
In order to widen the gap in the unit of $\sigma$, we need to 
increase the precision of the calculation of $B_K$ and $V_{cb}$
in lattice QCD.
In our calculation, one of the dominant source of error comes from the
matching factor for \bk ($\approx 4.4\%$) using the one-loop
perturbation theory.
Hence, it becomes essential to reduce the matching factor error.
The non-perturbative renormalization method (NPR) with the RI-MOM
\cite{Aoki:2007xm} can reduce this error down to the $\approx 2\%$
level.
In the previous work of Refs.~\cite{Kim:2013bta, Lytle:2013qoa}, the
NPR method has been applied to the staggered bilinear operators.
Here, we present preliminary results of the renormalization factors
of four-fermion operators relevant to \bk operator obtained using NPR
in the RI-MOM scheme with improved staggered fermions.
\section{Four-fermion operator renormalization in the RI-MOM scheme}
There are two kinds of color contraction of four-fermion operators. 
A general one-color trace four-fermion operator is defined as follows. 
\begin{align}
O_{\alpha, \I}(y) 
&= \sum_{\substack{A,B,\\C,D}}\sum_{\substack{c_{1},c_{2},\\c_{3},c_{4}}}
\big[
\ol{\chi}_{c_1}(y_A) \ol{(\gamma_{S_1} \otimes \xi_{F_1})}_{AB} \chi_{c_2}(y_B)
\big]
\big[ 
\ol{\chi}_{c_3}(y_C) \ol{(\gamma_{S_2} \otimes \xi_{F_2})}_{CD} \chi_{c_4}(y_D)
\big] 
\nonumber \\
& \qquad U_{AD;c_{1}c_{4}}(y)U_{BC;c_{2}c_{3}}(y) \,,
\end{align}
and a general two-color trace four-fermion operator is defined as follows. 
\begin{align}
O_{\alpha, \II}(y) 
&= \sum_{\substack{A,B,\\C,D}}\sum_{\substack{c_{1},c_{2},\\c_{3},c_{4}}}
\big[
\ol{\chi}_{c_1}(y_A) \ol{(\gamma_{S_1} \otimes \xi_{F_1})}_{AB} \chi_{c_2}(y_B)
\big]
\big[ 
\ol{\chi}_{c_3}(y_C) \ol{(\gamma_{S_2} \otimes \xi_{F_2})}_{CD} \chi_{c_4}(y_D)
\big] 
\nonumber \\
& \qquad U_{AB;c_{1}c_{2}}(y)U_{CD;c_{3}c_{4}}(y)\,,
\end{align}
where $\alpha = [S_1 \otimes F_1][S_2 \otimes F_2]$ is an operator
index, and $c_i$ are color indices.
The $y$ represents a coordinate of the hypercube with its lattice spacing $2a$.
The indices $A$, $B$, $C$ and $D$ are hypercubic vectors: for example,
$A = (1,1,0,0)$.
Here, we use the notation of $y_A = 2y + A$. 
$U_{AB;c_1 c_2}(y)$ is a gauge link, an average of the shortest paths
which connect $y_A$ and $y_B$ as products of HYP-smeared fat links.
$\gamma_{S}$ represents the spin and $\xi_{F}$ the taste.
Here, $\chi(y_B)$ represents HYP-smeared staggered quark field.
We calculate the amputated Green's function using same method
introduced in Ref.~\cite{Kim:2013bta}.

As an example, we choose the four fermion operators used to calculate
$B_K$ in order to illustrate how the NPR method produces the matching
factors.
We introduce the following simple notations for the operators.
\begin{align}
& O_{1} \equiv O_{[V \otimes P][V \otimes P], \I} \,, \qquad
O_{2} \equiv O_{[V \otimes P][V \otimes P], \II}\,,
\nonumber \\
& O_{3} \equiv O_{[A \otimes P][A \otimes P], \I} \,, \qquad
O_{4} \equiv O_{[A \otimes P][A \otimes P], \II} \,.
\label{eq:op}
\end{align}
First, we divide the lattice operators into two classes: (C) the
diagonal operators defined in Eq.\eqref{eq:op}, \{$O_1$, $O_2$, $O_3$, $O_4$\},
which have the $\xi_{5}$ tastes in both bilinears, and (D) the
off-diagonal operators which are remaining operators with taste
different from $\xi_{5}$.
The tree level \bk operator is sum of the operators in the (C) class.
\begin{align}
O^{\tree}_{B_K} = 
O^{\tree}_{1} + 
O^{\tree}_{2} + 
O^{\tree}_{3} + 
O^{\tree}_{4}
\end{align}
The projection operators are also defined in the same way in Eq.\eqref{eq:op}
as follows.
\begin{align}
\mathbb{P}_{1} &\equiv 
\frac{1}{N} 
\ol{\ol{(\gamma_{\mu}^{\dagger} \otimes \xi_{5}^{\dagger})}}_{BA} 
\ol{\ol{(\gamma_{\mu}^{\dagger} \otimes \xi_{5}^{\dagger})}}_{DC}
\delta_{c_4 c_1}\delta_{c_3 c_2} \,,&
\mathbb{P}_{2} &\equiv 
\frac{1}{N} 
\ol{\ol{(\gamma_{\mu}^{\dagger} \otimes \xi_{5}^{\dagger})}}_{BA} 
\ol{\ol{(\gamma_{\mu}^{\dagger} \otimes \xi_{5}^{\dagger})}}_{DC}
\delta_{c_2 c_1}\delta_{c_4 c_3} \,, 
\nonumber\\
\mathbb{P}_{3} &\equiv 
\frac{1}{N} 
\ol{\ol{(\gamma_{\mu 5}^{\dagger} \otimes \xi_{5}^{\dagger})}}_{BA} 
\ol{\ol{(\gamma_{\mu 5}^{\dagger} \otimes \xi_{5}^{\dagger})}}_{DC}
\delta_{c_4 c_1}\delta_{c_3 c_2} \,,& 
\mathbb{P}_{4} &\equiv 
\frac{1}{N} 
\ol{\ol{(\gamma_{\mu 5}^{\dagger} \otimes \xi_{5}^{\dagger})}}_{BA} 
\ol{\ol{(\gamma_{\mu 5}^{\dagger} \otimes \xi_{5}^{\dagger})}}_{DC}
\delta_{c_2 c_1}\delta_{c_4 c_3} \,.
\end{align}
Here, $N$ is normalization factor given as follows.
\begin{align}
N = 3072 = 
\underbrace{4^4}_{\text{spin}} 
\times 
\underbrace{4^4}_{\text{taste}}
\times 
(
\underbrace{3}_{\text{1-color trace}} + 
\underbrace{9}_{\text{2-color trace}}
)
\end{align}
We fix the normalization factor $N$ such that, when we apply the
projection operators to the tree level amputated Green's function, it
satisfies the following conditions.
\begin{align}
\label{eq:tree:1}
&\tr[\Lambda^{\tree}_{B_K} \mathbb{P}_{1}] 
= \tr[\Lambda^{\tree}_{B_K} \mathbb{P}_{2}] 
=\tr[\Lambda^{\tree}_{B_K} \mathbb{P}_{3}] 
=\tr[\Lambda^{\tree}_{B_K} \mathbb{P}_{4}] 
=1 \\
&\tr[\Lambda^{\tree}_{B_K} \mathbb{P}_{(D)}]
=\tr[\Lambda^{\tree}_{(D)} \mathbb{P}_{(C)}] 
=0
\label{eq:tree:2}
\end{align}
Here, note that the diagonal terms equal to one and the off-diagonal
terms becomes zero.

The renormalized \bk operator is defined as follows.
\begin{align}
O^{R}_{B_K} 
= z_{1}O^{B}_{1} + z_{2}O^{B}_{2} 
+ z_{3}O^{B}_{3} + z_{4}O^{B}_{4} 
+ \sum_{\alpha \in (D)} z_{\alpha}O^{B}_{\alpha}\,,
\end{align}
where the superscript R (B) denotes renormalized (bare) quantity, and
the coefficients $z_i$ are renormalization factors.
The renormalization of quark fields is defined as follows.
\begin{align}
\chi^{R} = Z^{1/2}_{q} \chi^{B}
\end{align}
The amputated Green's function is obtained by multiplying the inverse
propagators to the unamputated Green's function. 
Hence the renormalized amputated Green's function is as follows.
\begin{align}
\label{eq:renorm}
\Lambda^{R}_{B_K} 
= \frac{z_1}{z_q^2}\Lambda^{B}_{1} 
+ \frac{z_2}{z_q^2}\Lambda^{B}_{2} 
+ \frac{z_3}{z_q^2}\Lambda^{B}_{3} 
+ \frac{z_4}{z_q^2}\Lambda^{B}_{4}
+ \sum_{\alpha \in (D)} \frac{z_{\alpha}}{z^2_q}\Lambda^{B}_{\alpha}
\end{align}
The RI-MOM scheme prescription is that the renormalized quantity is
equal to its tree level value.
\begin{align}
\label{eq:RI-MOM}
\tr[\Lambda^{R}_\alpha(\wtd{p},\wtd{p},\wtd{p},\wtd{p})
  \mathbb{P}_\beta] 
& = \tr[\Lambda^{\tree}_\alpha(\wtd{p},\wtd{p},\wtd{p},\wtd{p})
  \mathbb{P}_\beta] \,,
\end{align}
where $\wtd{p}$ is a momentum defined in the reduced Brillouin
zone.\footnote{Please refer to Ref.~\cite{Kim:2013bta} for more
  details.}
We define the projected amputated Green's function as follows.
\begin{align}
\Gamma^{B}_{\alpha \beta} &\equiv 
\frac{1}{z^2_q}\tr[\Lambda^{B}_{\alpha}\mathbb{P}_{\beta}]
\end{align}
Hence, from Eq.~\eqref{eq:tree:1}, Eq.~\eqref{eq:tree:2},
Eq.~\eqref{eq:renorm} and Eq.~\eqref{eq:RI-MOM}, we obtain
the following relations.
\begin{align}
1 &= 
z_{1}\Gamma^{B}_{1\alpha} + 
z_{2}\Gamma^{B}_{2\alpha} + 
z_{3}\Gamma^{B}_{3\alpha} + 
z_{4}\Gamma^{B}_{4\alpha} + 
\sum_{\gamma \in (D)} 
z_{\gamma}\Gamma^{B}_{\gamma \alpha} 
\,, \quad 
\alpha \in (C) \\
0 &= 
z_{1}\Gamma^{B}_{1\beta} + 
z_{2}\Gamma^{B}_{2\beta} + 
z_{3}\Gamma^{B}_{3\beta} + 
z_{4}\Gamma^{B}_{4\beta} + 
\sum_{\gamma \in (D)} 
z_{\gamma}\Gamma^{B}_{\gamma \beta} 
\,, \quad 
\beta \in (D)
\end{align}
%
%
We can express these equations as a matrix equation.
\begin{align}
\vec{z}_{\tree} = \vec{z} \cdot \hat{\Gamma}^{B} \,, 
\end{align}
where $\vec{z}_{\tree}$ and $\vec{z}$ are vectors as follows.
\begin{align}
&\vec{z}_{\tree} = (1, 1, 1, 1, 0, \cdots, 0)\,, \quad 
\vec{z} = (z_{1}, z_{2}, z_{3}, z_{4}, z_5, z_6, \cdots)
\end{align}
where $z_i$ with $i\geq 5$ are renormalization factors of off-diagonal
operators.
The $\hat{\Gamma}^{B}$ is a matrix as follows. The upper-left (red) block
elements are diagonal terms and others are off-diagonal terms.
\[
\hat{\Gamma}^{B} = \left(
\begin{array}{c  c  c  c :c c c}
\textcolor{red}{\Gamma^{B}_{11}} & 
\textcolor{red}{\Gamma^{B}_{12}} & 
\textcolor{red}{\Gamma^{B}_{13}} & 
\textcolor{red}{\Gamma^{B}_{14}} & 
\Gamma^{B}_{15} & 
\Gamma^{B}_{16} & 
\cdots \\
\textcolor{red}{\Gamma^{B}_{21}} & 
\textcolor{red}{\Gamma^{B}_{22}} & 
\textcolor{red}{\Gamma^{B}_{23}} & 
\textcolor{red}{\Gamma^{B}_{24}} & 
\Gamma^{B}_{25} & 
\Gamma^{B}_{26} & 
\cdots \\
\textcolor{red}{\Gamma^{B}_{31}} & 
\textcolor{red}{\Gamma^{B}_{32}} & 
\textcolor{red}{\Gamma^{B}_{33}} & 
\textcolor{red}{\Gamma^{B}_{34}} & 
\Gamma^{B}_{35} & 
\Gamma^{B}_{36} & 
\cdots \\
\textcolor{red}{\Gamma^{B}_{41}} & 
\textcolor{red}{\Gamma^{B}_{42}} & 
\textcolor{red}{\Gamma^{B}_{43}} & 
\textcolor{red}{\Gamma^{B}_{44}} & 
\Gamma^{B}_{45} & 
\Gamma^{B}_{46} & 
\cdots \\
\hdashline
\Gamma^{B}_{51} & 
\Gamma^{B}_{52} & 
\Gamma^{B}_{53} & 
\Gamma^{B}_{54} & 
\Gamma^{B}_{55} & 
\Gamma^{B}_{56} & 
\cdots \\
\Gamma^{B}_{61} & 
\Gamma^{B}_{62} & 
\Gamma^{B}_{63} & 
\Gamma^{B}_{64} & 
\Gamma^{B}_{65} & 
\Gamma^{B}_{66} & 
\cdots \\
\vdots & \vdots & \vdots & \vdots & \vdots & \vdots & \ddots
\end{array}
\right) \]
Hence, we can compute $z$-factors from the inverse of
$\hat{\Gamma}^{B}$ matrix as follows.
\begin{align}
\vec{z} = \vec{z}_{\tree} \cdot (\hat{\Gamma}^{B})^{-1}
\end{align}
The $\hat{\Gamma}^{B}$ can be rewritten by sub-matrices as follows.
\begin{align}
\hat{\Gamma}^B =
\begin{pmatrix}
X_{4 \times 4} & Y_{4 \times 20} \\
Z_{20 \times 4} & W_{20 \times 20}
\end{pmatrix}
\end{align}
Here, $X$ is diagonal terms, and $Y$ is off-diagonal terms.
\begin{align}
X =
\begin{pmatrix}
\Gamma^{B}_{11} & \Gamma^{B}_{12} & \Gamma^{B}_{13} & \Gamma^{B}_{14} \\
\Gamma^{B}_{21} & \Gamma^{B}_{22} & \Gamma^{B}_{23} & \Gamma^{B}_{24} \\
\Gamma^{B}_{31} & \Gamma^{B}_{32} & \Gamma^{B}_{33} & \Gamma^{B}_{34} \\
\Gamma^{B}_{41} & \Gamma^{B}_{42} & \Gamma^{B}_{43} & \Gamma^{B}_{44}
\end{pmatrix} \,,
\qquad
Y =
\begin{pmatrix}
\Gamma^{B}_{15} & \Gamma^{B}_{16} & \cdots \\
\Gamma^{B}_{25} & \Gamma^{B}_{26} & \cdots \\
\Gamma^{B}_{35} & \Gamma^{B}_{36} & \cdots \\
\Gamma^{B}_{45} & \Gamma^{B}_{46} & \cdots
\end{pmatrix}
\end{align}
The number of the off-diagonal operators is 20 
and they are $\{O_5, O_6, \cdots\} = 
\{
(S \otimes V)(S \otimes V)_{\I}, 
(S \otimes V)(S \otimes V)_{\II}, 
(S \otimes A)(S \otimes A)_{\I}, 
\cdots\}$.
We assume that $Z \simeq Y^{T} \simeq
\mathcal{O}(\alpha_s)$\footnote{Note that this approximation is good
  within factor of 3.} and $W \simeq 1 + \mathcal{O}(\alpha_s)$.
The inverse of block matrix is 
\begin{align}
&(\hat{\Gamma}^B)^{-1} =
\begin{pmatrix}
(X - Y W^{-1} Z)^{-1}
& - X^{-1} Y (W - Z X^{-1} Y)^{-1} \\
- W^{-1} Z (X - Y W^{-1} Z)^{-1}
& (W - Z X^{-1} Y)^{-1}
\end{pmatrix} \,.
\end{align}
Using power series expansion in $Y$ and $Z$, it becomes as follows. 
\begin{align}
(\hat{\Gamma}^B)^{-1} &\simeq
\begin{pmatrix}
X^{-1} + X^{-1} Y W^{-1} Z X^{-1} &
- X^{-1} Y (W^{-1} + W^{-1} Z X^{-1} Y W^{-1}) \\
- W^{-1} Z (X^{-1} + X^{-1} Y W^{-1} Z X^{-1}) & 
W^{-1} + W^{-1} Z X^{-1} Y W^{-1}
\end{pmatrix}
\end{align}
With our assumption, $Z \simeq Y^{T}$ and $W \simeq 1$,
\begin{align}     
(\hat{\Gamma}^B)^{-1} &\simeq
\begin{pmatrix}
X^{-1} + X^{-1} Y  Y^{T} X^{-1} & 
- X^{-1} Y (1 + Y^T X^{-1} Y ) \\
- Y^T (X^{-1} + X^{-1} Y Y^T X^{-1}) & 
1 + Y^T X^{-1} Y
\end{pmatrix} \,.
\end{align}

\section{Results}
\begin{figure}[htpb]
\subfigure[m-fit]{
\label{fig:m-fit}
\includegraphics[width=0.49\textwidth]{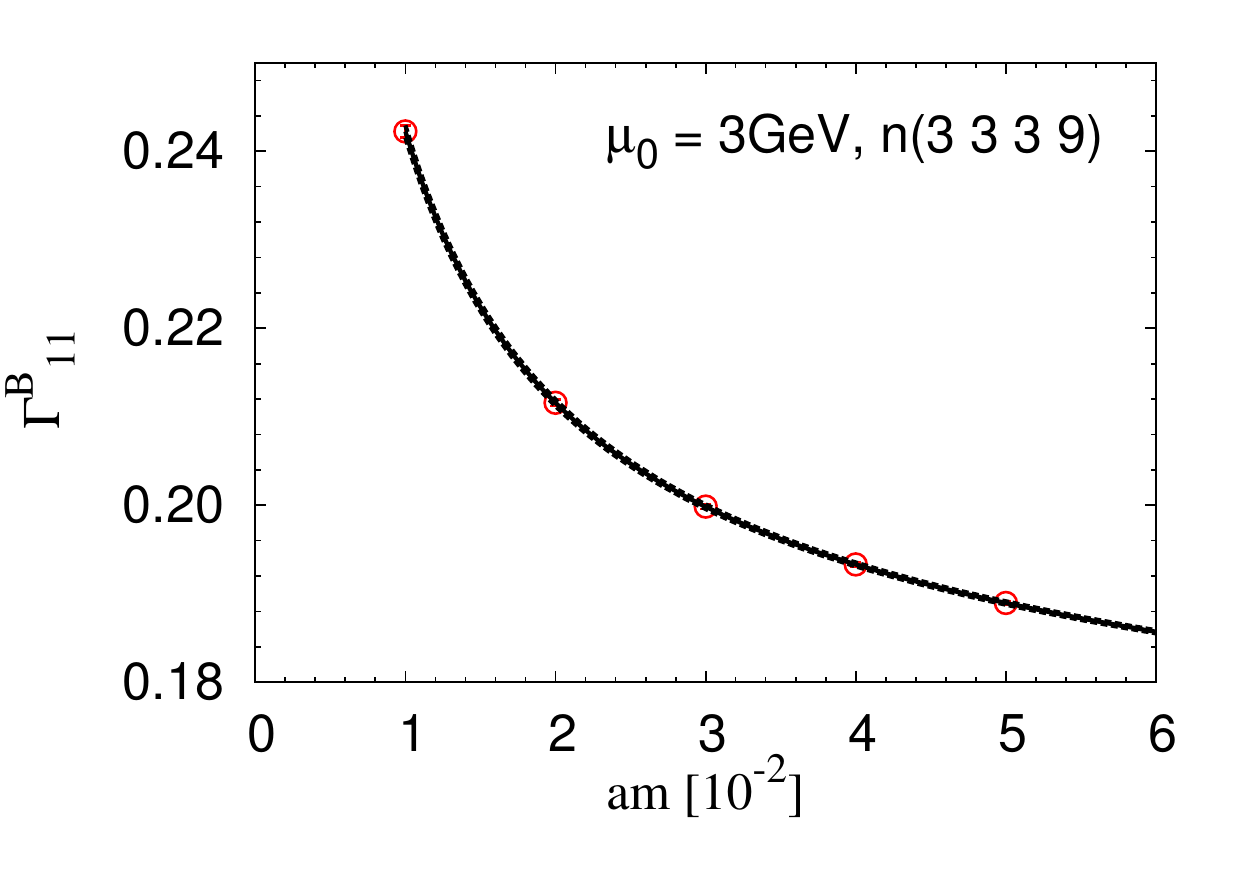}
}
\subfigure[p-fit]{
\label{fig:p-fit}
\includegraphics[width=0.49\textwidth]{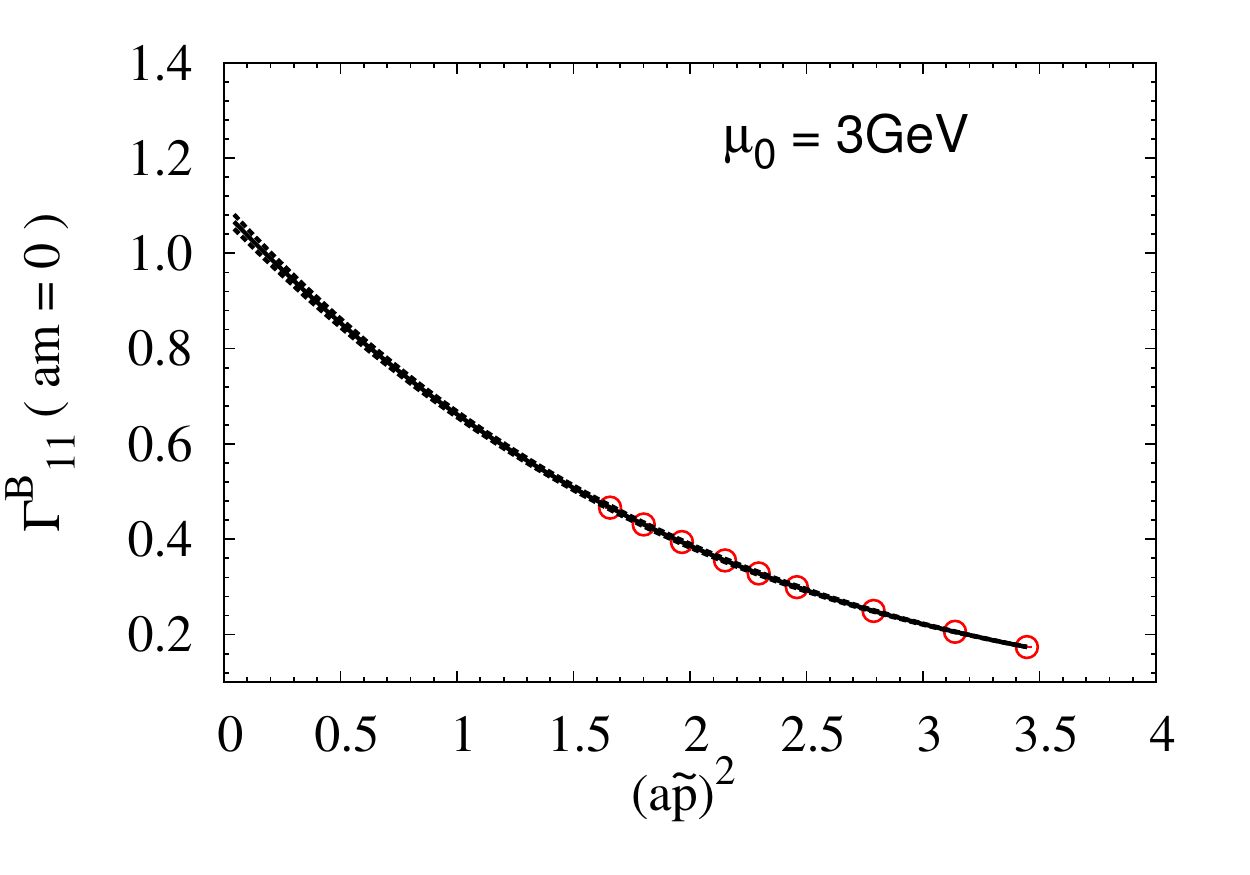}
}
\caption{m-fit and p-fit plot at $\mu_0 = 3\GeV$}
\end{figure}

First, we present the data analysis of diagonal terms.
Let us consider the element $\Gamma^{B}_{11}$ of $\hat{\Gamma}^B$ matrix.
We measure the data for 5 valence quark masses and 9 external momenta.
The scale of raw data is determined by external momentum ($\mu =
|\wtd{p}|$).
Hence, we convert the scale of raw data to the common scale $\mu_{0}=
3\GeV$ using two-loop RG evolution \cite{Kim:2014tda}.
We fit the data with respect to quark mass for a fixed momentum to the
following function $f$ suggested from Ref.~\cite{Blum:2001sr}
based on the Weinberg theorem \cite{Weinberg:1959nj}.
\begin{align}
f(m, a, \wtd{p}) 
= c_{1} 
+ c_{2} \cdot (am) 
+ c_{3} \cdot \frac{1}{(am)} 
+ c_{4} \cdot \frac{1}{(am)^2},
\end{align}
where $m$ is a valence quark mass. 
After m-fit, we take the $c_1(a\wtd{p})$ as chiral limit
values.
Because of the sea quark determinant contributions ($c_{3}, c_{4}
\propto (m^2_{\ell}m_s)^\nu$) with $\nu$ the number of zero modes,
these pole terms contributions vanish in the chiral limit.
The fitting results are presented in Table~\ref{tab:m-fit} and
Fig.~\ref{fig:m-fit}.
\begin{table}[htbp]
\center
\begin{tabular}{c || c | c | c | c | c }
\hline
\hline
$\mu_0$ & $c_1$ & $c_2$ & $c_3$ & $c_4$ & $\chi^2/\text{dof}$ \\
\hline
3GeV & 0.17991(20) & -0.0975(15) & 0.0007118(85) & -0.000000790(36) & 0.00194(40) \\
\hline
\hline
\end{tabular}
\caption{
\label{tab:m-fit}
m-fit
}
\end{table}
We fit $c_1(a\wtd{p})$ to the following fitting function.
\begin{align}
g(a\wtd{p}) &= 
b_{1} + 
b_{2} \cdot (a\wtd{p})^2 + 
b_{3} \cdot ((a\wtd{p})^2)^2 + 
b_{4} \cdot (a\wtd{p})^4 + 
b_{5} \cdot ((a\wtd{p})^2)^3
\end{align}
To avoid non-perturbative effects at small $(a \wtd{p} )^2 \leq 1$,
we choose the momentum window as $(a \wtd{p} )^2 > 1$.
Because we assume that those terms of
$\mathcal{O}((a\wtd{p})^2)$ and higher order are pure lattice
artifacts, we take the $b_1$ as $\Gamma^{B}_{11}$ value at $\mu_0=3\GeV$
in the RI-MOM scheme.
The fitting result and plot are presented in Table~\ref{tab:p-fit} and
Fig.~\ref{fig:p-fit}.
\begin{table}[htbp]
\center
\begin{tabular}{l || l | l | l | l | l | l }
\hline
\hline
$\mu_0$ & $b_1$ & $b_2$ & $b_3$ & $b_4$ & $b_{5}$ & $\chi^2/\text{dof}$ \\
\hline
3GeV & 1.088(16) & -0.515(18) & 0.0953(74) & 0.0020(65) & -0.00663(91) & 0.08(17) \\
\hline
\hline
\end{tabular}
\caption{
\label{tab:p-fit}
p-fit
}
\end{table}
Similarly, we analyse the whole elements of $\hat{\Gamma}^{B}$ matrix.
Results of diagonal terms in the inverse of $\hat{\Gamma}^{B}$ are
\begin{align}
X^{-1}&=
\begin{pmatrix}
 1.333(32) & -0.793(45) &  0.336(21) &  0.007(31)  \\
-0.726(44) &  1.940(41) & -0.008(30) & -0.047(33)  \\
 0.341(30) &  0.032(45) &  1.222(32) & -0.604(37)  \\
 0.018(45) & -0.084(52) & -0.637(39) &  1.543(36) 
\end{pmatrix} \\
X^{-1} Y  Y^{T} X^{-1}
&=\begin{pmatrix}
 0.0127(15)  & -0.0079(11)  &  0.0020(17)  & -0.0001(10)  \\
-0.00713(92) &  0.0064(11)  & -0.0010(11)  & -0.00045(73) \\
 0.0020(18)  &  0.0000(15)  &  0.0200(90)  & -0.0103(48)  \\
 0.0001(12)  & -0.0012(11)  & -0.0109(52)  &  0.0059(28)  
\end{pmatrix}
\end{align}
We obtain $\vec{z}$ in RI-MOM scheme at $\mu_0=3\GeV$.
We convert the scheme from RI-MOM to $\MSb$ using two-loop RG
evolution.
Results are summarized in Table \ref{tab:rg-evol}.
\begin{table}[htbp]
\center
\begin{tabular}{c || c | c}
\hline
\hline
& RI-MOM(3GeV) & $\MSb$(3GeV) \\
\hline
$z_1$ & 0.9666(78) & 0.9812(79) \\
$z_2$ & 1.095(30)  & 1.111(31)  \\
$z_3$ & 0.9139(73) & 0.9277(74) \\
$z_4$ & 0.898(28)  & 0.912(29)  \\
\hline
\hline
\end{tabular}
\caption{The $z$-factors in RI-MOM and $\MSb$ scheme. Here, the errors
  are purely statistical.}
\label{tab:rg-evol}
\end{table}

Now let us switch the gear to the systematic errors.
The first systematic error comes from the diagonal correction terms,    
$X^{-1} Y W^{-1} Z X^{-1} \approx X^{-1} Y Y^{T} X^{-1}$.
We quote $E_{diag} \equiv \vec{z}_\tree \cdot  X^{-1} Y Y^{T} X^{-1}$
as this error.
The second systematic error comes from the off-diagonal correction
terms.
Their size ($- X^{-1} Y$) are typically less than $7\%$.
However, thanks to the wrong taste suppression ($\ll 1\%$)
\cite{Wlee:thesis}, their effect becomes $\ll 0.07\%$.
Hence we neglect them without loss of generality.
Another systematic error comes from truncated higher order of the
two-loop RG evolution factor (RI-MOM$\to\MSb$).
We quote $E_{t} \equiv z_{i} \cdot \alpha_s^3$ as this error.
\begin{table}[htbp]
\center
\begin{tabular}{c || c | c | c | c}
\hline
\hline
& $\MSb$(3GeV) & $E_{diag}$ & $E_{t}$ & $E_{tot}$ \\
\hline
$z_1$ &  0.9812(79) & 0.0077 & 0.0144 & 0.0163 \\
$z_2$ &  1.111(31)  & 0.0027 & 0.0163 & 0.0165 \\
$z_3$ &  0.9277(74) & 0.0101 & 0.0136 & 0.0171 \\
$z_4$ &  0.912(29)  & 0.0050 & 0.0134 & 0.0143 \\
\hline
\hline
\end{tabular}
\caption{The systematic errors of $z$-factors in $\MSb$ scheme at
  $3\GeV$.  $E_{tot}$ represents the total systematic error.}
\label{tab:sys-err}
\end{table}
We add these systematic errors ($E_{diag}$ and $E_t$) in quadrature as
summarized in Table \ref{tab:sys-err}.
%


In addition, we compare the NPR result($\MSb$ [NDR]) with those of
one-loop perturbative matching.
We quote truncated two-loop uncertainty: $E_{t}^{\text{one-loop}}
\equiv z_{i} \cdot \alpha_{s}^2$ as our estimate of the systematic
error of one-loop matching.
\begin{table}[htbp]
\center
\begin{tabular}{c || c | c | c}
\hline
\hline
& NPR(3GeV)  & one-loop(3GeV)  & $\Delta$\\
\hline
$z_1$ &  0.981(8)(16) & 1.035(62)  & 0.83 $\sigma$ \\
$z_2$ &  1.111(31)(17)   & 1.120(67)  & 0.12 $\sigma$ \\
$z_3$ &  0.928(7)(17) & 1.043(63)  & 1.75 $\sigma$ \\
$z_4$ &  0.912(29)(14)   & 0.953(57)  & 0.63 $\sigma$ \\
\hline
\hline
\end{tabular}
\caption{Comparison NPR results and one-loop perturbative matching factors at $3\GeV$.
}
\end{table}
Here, note that the results of NPR are consistent with those of
one-loop matching within $2\sigma$.
This indicates that our NPR results are quite reasonable.
\section{Acknowledgments}
The research of W.~Lee is supported by the Creative Research
Initiatives Program (No.~2014001852) of the NRF grant funded by the
Korean government (MEST).
W.~Lee would like to acknowledge the support from KISTI supercomputing
center through the strategic support program for the supercomputing
application research (KSC-2013-G2-005).
%
\bibliographystyle{JHEP}
\bibliography{ref}

\end{document}